\documentclass[a4paper]{report}
\usepackage[utf8]{inputenc}
\usepackage[T1]{fontenc}
\usepackage{float}
\usepackage[table]{xcolor}
\usepackage{RJournal}
\usepackage{amsmath,amssymb,array}
\usepackage{booktabs}
\usepackage{soul} 

\newcommand{\grb}[1]{\cellcolor[gray]{0.8}{#1}}

\def\btheta{\mbox{\boldmath $\theta$}}
\def\bpsi{\mbox{\boldmath $\psi$}}

\def\bkappa{\mbox{\boldmath $\kappa$}}

\def\bs{\mathbf{s}}
\def\bA{\mathbf{A}}
\def\bB{\mathbf{B}}

\def\qmo{``}
\def\qmc{''}

\begin{document}

\sectionhead{Contributed research article}
\volume{XX}
\volnumber{YY}
\year{20ZZ}
\month{AAAA}

\begin{article}
\title{Value--at--Risk Prediction in R with the GAS Package}
\author{by David Ardia, Kris Boudt and Leopoldo Catania}

\maketitle

\abstract{
GAS models have been recently proposed in time--series econometrics as valuable tools for signal extraction and prediction. This paper details how financial risk managers can use GAS models for Value--at--Risk (VaR) prediction using the novel \CRANpkg{GAS} package for \code{R}. Details and code snippets for prediction, comparison and backtesting with GAS models are presented. An empirical application considering Dow Jones Index constituents investigates the VaR forecasting performance of GAS models.}

\section{Introduction}

Following the regulatory process of the Basel Accords (currently the Basel III Accords), banks and financial institutions are required to meet capital requirements, and must rely on state--of--the--art risk systems. In particular, they must assess the uncertainty about the future values of their portfolios, and estimate the extent and the likelihood of potential losses. This is typically achieved in two steps. First, the distribution of future profit \& loss (\emph{i.e.}, future portfolios' or assets' returns) is modelled. Second, financial risk is measured from the distribution; nowadays, the Value--at--Risk (VaR) risk measure is the standard \citep{jorion.1997}. This metric gives, for a given time horizon, the asset's loss (or return) that is expected to be exceeded with a given probability level $\alpha$ (referred to as the \emph{risk level}, and which is typically set to one or five percent, \emph{i.e.}, $\alpha\in\{0.01,0.05\}$). Hence, the VaR is nothing else than a given percentile of the returns distribution. The popularity of VaR mostly relies on: (i) the simple rationale behind it, (ii) the ease of computation, and (iii) its role in the financial regulation \citep[see][]{basel.2010}.

Formally, assuming a continuous cumulative density function (\emph{cdf}) with time--varying parameters $\btheta_t\in\mathbb{R}^d$ and additional static parameters $\bpsi\in\mathrm{R}^q$, $F(\cdot;\btheta_t, \bpsi)$, for the asset log--return at time $t$, $r_t\in\mathbb{R}$, the $VaR_t(\alpha)$ is given by:
\begin{align*}
VaR_t(\alpha) \equiv F^{-1}(\alpha; \btheta_t, \bpsi) \,,
\end{align*}
where $F^{-1}(\cdot)$ denotes the inverse of the \emph{cdf}, \emph{i.e.}, the quantile function. It follows that $VaR_t(\alpha)$ is nothing more than the $\alpha$--quantile of the return distribution at time $t$.\footnote{Sometimes VaR is defined with respect to the the loss variable $l_t = -r_t$, \emph{i.e.}, the negative of the return. Clearly, all the arguments of this paper can be easily adapted to this case.} Evidently, remaining within the fully parametric framework considered here, the crucial point is the determination of $F\left(\cdot\right)$ and its parameters $\btheta_t$ and $\bpsi$. Indeed, the majority of the financial econometrics literature focuses on this aspect; see \emph{e.g.}, \cite{mcneil.2015}.

Recently, the new class of Score Driven (SD) models has been introduced by \citet{creal_etal.2013} and \citet{harvey.2013} offering an alternative to the Generalized AutoRegressive Heteroscedasticity (GARCH) framework pioneered by \citet{bollerslev.1986} to model the conditional variance of financial returns. SD models are also referred to as Generalized Autoregressive Score (GAS) and Dynamic Conditional Score (DCS) models. In this paper, we follow the GAS nomenclature.

Formally, in GAS models the vector of time--varying parameters, $\btheta_t$, is updated through a dynamic equation based on the score of the conditional\footnote{The conditioning is intended with respect to the past observations $r_{t-s}$ $(s>0)$, however, for notational purposes, this is not always reported.} probability density function of $r_t$, $f(\cdot;\btheta_t,\bpsi)$, \emph{i.e.}:
\begin{align}\label{eq:update}
\btheta_{t+1} \equiv \bkappa + \bA\bs_t + \bB\btheta_t \,,
\end{align}
where $\bs_t$ is the scaled score of $f(\cdot;\btheta_t,\bpsi)$ with respect to $\btheta_t$, evaluated in $r_t$; see \citet{creal_etal.2013}. The coefficients $\bkappa$, $\bA$ and $\bB$ control for the evolution of $\btheta_t$ and need to be estimated along with $\bpsi$ from the data, usually by Maximum Likelihood. GAS models have been employed for a variety of applications in financial econometrics, mostly because they offer a link between the two most common frameworks to model volatility, namely: GARCH models and stochastic volatility (SV) models \citep{taylor.1986}. Indeed, while resorting on straightforward estimation procedures as for GARCH, they update models parameters accounting for the whole shape of the conditional distribution of the data, as in the SV context; see \citet{koopman_etal.2016}. Hence, they offer a good trade--of between ease of estimation and flexibility.

Within the \code{R} statistical environment, the \CRANpkg{GAS} package of \citet{GAS} allows practitioners and researchers to easily deal with GAS models. \CRANpkg{GAS} implements routines to estimate, predict, and simulate with GAS models; see \citet{ardia_etal.2016}.

The aim of this paper is to show how the \CRANpkg{GAS} package can be used for VaR evaluation, prediction and backtesting. We do this by detailing the functionality of the \CRANpkg{GAS} package devoted to these aspects as well as reporting an empirical application using financial time series. We focus on the three major steps a practitioner involved in risk management faces during his job: (i) prediction of future VaR levels, (ii) backtesting, and (iii) comparison with alternative models. The empirical part of the article deals with these three points from an applied perspective while the computational part details the \CRANpkg{GAS} functionalities devoted to VaR.

The article is organized in the following manner. We first introduce a general GAS model suited for financial returns. Then we review the methodology for assessing and comparing VaR predictions in the context of GAS models. \CRANpkg{GAS} functionalities devoted to VaR analysis are subsequently illustrated. Finally, we report an empirical application using a moderately large set of financial returns.

\section{A skew--Student--t GAS model with time--varying volatility}

Financial returns exhibit several stylized facts that need to be taken into consideration in order to produce reliable risk forecasts. Empirically, the distribution of returns is (left--)skewed and fat--tailed, and its variance is time--varying (\emph{i.e.}, returns exhibit the so--called \emph{volatility clustering}); see, \emph{e.g.}, \citet{mcneil.2015}.
	
To account for these features, we consider a very flexible specification, in which we assume that the log--return at time $t$, $r_t$, is distributed conditionally on past observations as follows:
\begin{align*}
r_{t}\vert\mathcal{I}_{t-1}\sim\mathcal{SKST}(r_t;\mu,\sigma_{t},\xi,\nu) \,,
\end{align*}
where $\mathcal{I}_{t-1}$ is the information set up to time $t-1$, and $\mathcal{SKST}\left(r_t;\cdot\right)$ denotes the skew--Student--t distribution of \citet{fernandez_steel.1998} with location $\mu\in\mathbb{R}$, time--varying scale $\sigma_t>0$, and shape and skewness parameters $\nu>2$ and $\xi>0$, respectively. We parametrise the $\mathcal{SKST}$ distribution as in \citet{bauwens_laurent.2005} such that $\mathbb{E}_{t-1}\left[r_t\right]=\mu$ and $\mathrm{Var}_{t-1}\left[r_t\right] = \sigma_t^2$. The log--density of the $\mathcal{SKST}$ distribution evaluated in $r_t$ is given by:
\begin{align*}
\log f_{\mathcal{SKST}}\left(r_t;\mu,\sigma_t,\xi,\nu\right) = \log g + \log s + c - \frac{\nu+1}{2}\log\left[1 + \frac{\left[\left(\frac{r_t - \mu}{\sigma_t}\right)s + m\right]^2}{\left(\nu-2\right)\left(\xi_t^*\right)^2}\right] \,,
\end{align*}
where:
\begin{align*}
m &\equiv \mu_1\left(\xi - \frac{1}{\xi}\right) \\
s &\equiv \sqrt{\left(1 - \mu_1^2\right)\left(\xi^2 + \frac{1}{\xi^2}\right) + 2\mu_1^2 - 1}\\
g &\equiv \frac{2}{\xi + \frac{1}{\xi}}\\
c &\equiv \frac{1}{2}\left[\log\nu - \log\left(\nu -2\right) - \log\pi + \log\nu\right] + \log\Gamma\left(\frac{\nu + 1}{2}\right) - \log\Gamma\left(\frac{\nu}{2}\right) \,,
\end{align*}
with:
\begin{align*}
\mu_1 \equiv \frac{2\sqrt{\nu - 2}}{(\nu - 1)}\frac{\Gamma(\frac{\nu+1}{2})}{\Gamma(\tfrac{\nu}{2})\Gamma(\tfrac{1}{2})} \,,
\end{align*}
and:
\begin{align*}
\xi_t^* \equiv \begin{cases}
1 & \mbox{if } z_t = 0 \\
\frac{1}{\xi} & \mbox{if } z_t<0 \\
\xi & \mbox{if } z_t>0
\end{cases} \,,
\end{align*}
where $z_t \equiv \left(\frac{r_t - \mu}{\sigma_t}\right)s + m$. It follows that, similar to \citet{laurent_etal.2016}, the time--varying GAS parameter $\btheta_t$ in~\eqref{eq:update} is given by $\btheta_t \equiv \theta_t \equiv \log\sigma_t$ and that its time--variation is also determined by the static parameters $\bpsi \equiv (\mu,\nu,\xi)^\prime$. The score of $\mathcal{SKST}(r_t;\cdot)$ with respect to $\theta_t$, $s_t$, is given by:
\begin{align*}
s_t \equiv \left(\frac{sz_t\left(\nu + 1\right)\left(r_t-\mu\right)}{\left(\xi_t^*\right)^2\left(\nu - 2\right)} - 1\right) \,,
\end{align*}
and enters linearly in the updating equation~\eqref{eq:update}. Starting from the general $\mathcal{SKST}$ distribution, we recover as special cases:

\begin{itemize}
\item[(i)] The Student--t distribution, $\mathcal{ST}$, imposing $\xi = 1$;
\item[(ii)] The Normal distribution, $\mathcal{N}$, imposing $\nu = \infty$ and $\xi = 1$.
\end{itemize}

Given a series of $T$ log--returns, $r_1,\dots,r_T$, the model parameters are estimated by maximizing the log--likelihood function; see \citet{blasques_etal.2014b}. Prediction with GAS models is straightforward thanks to the recursive nature of the updating equation~\eqref{eq:update}. Specifically, the one--step ahead predictive distribution $F(\cdot;\hat\theta_{T+1},\bpsi)$ is available in closed form whereas the $h$--step ahead distribution ($h>1$) needs to be simulated; see \citet{blasques_etal.2016}. As a direct consequence, at time $t$, $VaR_{t+1}(\alpha)$ is directly available, whereas $VaR_{t+h}(\alpha)$ $(h>1)$ has to be evaluated as the quantile of simulated values.

\section{Backtesting and comparing the VaR}

The recursive method of forecasting \citep{marcellino_etal.2006} is usually employed in order to backtest the adequacy of a statistical model as well as to perform models comparison in terms of VaR predictions. The objective of a backtesting analysis is to verify the precision of the prediction by separating the estimation window and the evaluation period. Differently, models comparison usually order models according to a loss function. To this end, the full sample of $T$ returns is divided in an in--sample period of length $S$, and an out--of--sample period of length $H$. Model parameters are firstly estimated over the in--sample period, subsequently the $h$--step ahead prediction of the return distribution at time $S+h$ is generated along with the corresponding VaR level. These steps are repeated augmenting the in--sample period with new observations in a recursive way until we reach the end of the series, $T$. If during the data augmentation step, past observations are eliminated, we are considering a rolling window, otherwise we have an expanding window. In this paper, we use the rolling window configuration for our analysis and we set $h = 1$.

Once a series of VaR predictions is available, forecasts adequacy is assessed through backtesting procedures. VaR backtesting procedures usually checks the correct coverage of the unconditional and conditional left--tail of log--returns distribution. Correct unconditional coverage (UC) was first considered by \citet{kupiec.1995}, while correct conditional coverage (CC) by \citet{christoffersen.1998}. The main difference between UC and CC concerns the distribution we are focusing on. For instance, UC considers correct coverage of the left--tail of the marginal return distribution, $f(r_t)$, while CC deals with the conditional density $f(r_t\vert\mathcal{I}_{t-1})$. From an inferential perspective, UC looks at the ratio between the expected number of VaR violations implied by the chosen confidence level, $\alpha$, during the forecast period, \emph{i.e.}, $\alpha H$, and the realized VaR violations observed from the data. Formally, the Actual over Expected (AE) ratio is defined as:
\begin{align}\label{eq:AE}
AE \equiv \frac{\sum_{s = 1}^{H} d_{t+s}}{\alpha H} \,,
\end{align}
where $d_{t+s} \equiv \boldsymbol{1}_{\{r_{t+s}<VaR_{t+s}(\alpha)\}}$ and $\boldsymbol{1}_{\{ \cdot \}}$ is the indicator function equal to one if the condition holds, and zero otherwise. Hence, if $AE>1$ the model underestimate risk, while if $AE<1$ the model is too conservative. Clearly, both situations are problematic and imply capital losses, the former due to unexpected negative returns, the latter due to overestimation of capital requirements.

In order to investigate CC, \citet{christoffersen.1998} proposed a test on the series of VaR exceedance $\{d_{t+s},s = 1,\dots,H\}$, usually referred to as \qmo hitting series\qmc. Specifically, if correct conditional coverage is achieved by the model, VaR exceedances should be independently distributed over time.

Recently, a general testing procedure for dynamic quantile models, also suited for VaR backtesting, has been proposed by \citet{engle_manganelli.2004}. The dynamic quantile test (DQ) jointly tests for UC and CC and has more power than previous alternatives under some form of model misspecification. The series of interest is defined as $\{d_{t+s} - \alpha,s = 1,\dots,H\}$. Under correct model specification, we have the following moment conditions: $\mathbb{E}\left[d_{t+s} - \alpha\right] = 0$,   $\mathbb{E}_{t+s-1}\left[d_{t+s} - \alpha\right] = 0$,  $\mathbb{E}[(d_{t+s} - \alpha)(d_{t+j} - \alpha)] = 0$ for $j\neq s$; see \citet{engle_manganelli.2004}.

Real world applications generally consider several models for VaR prediction. If correct unconditional/conditional coverage is achieved by more than one model, the practitioner faces the problem of not being able to choose between different alternatives. In this situation, model comparison techniques are used in order to choose the best performing model. Models ranking is achieved thanks to the definition of a loss function. Among several available loss functions for quantile prediction \citep{mcaleer_daveiga.2008}, the Quantile Loss (QL) used for quantile regressions \citep{koenker_bassett.1978} is one the most frequent choice in the VaR context; see \citet{gonzalezriviera_etal.2004}. Formally, given a VaR prediction at confidence level $\alpha$ for time $t+1$, the associated quantile loss, $QL_{t+1}\left(\alpha\right)$, is defined as:
\begin{align*}
QL_{t+1}(\alpha) \equiv (\alpha - d_{t+1} )\left(y_{t+1} - VaR_{t+1}(\alpha\right)) \,.
\end{align*}
Evidently, QL is an asymmetric loss function that penalizes more heavily with weight $\left(1 - \alpha\right)$ the observations for which we observe returns VaR exceedance. Quantile losses are then averaged over the forecasting period, models with lower averages are preferred. Outperformance of model $\mathcal{A}$ versus model $\mathcal{B}$ is finally assessed looking at the ratio between the average QLs, associated with the two models, \emph{i.e.}, if $QL_\mathcal{A}/QL_\mathcal{B} < 1$ then model $\mathcal{A}$ outperforms model $\mathcal{B}$ and vice versa.

Summarizing, given a set of available models, a typical VaR forecasting exercise consists of the three major steps:

\begin{itemize}
\item[(i)] perform rolling forecast during the out--of--sample period,
\item[(ii)] statistical backtest of VaR predictions using UC, CC and DQ tests,
\item[(iii)] model comparison looking at the average QL of each model.
\end{itemize}

The next section is devoted to detail the implementation of these steps with the \CRANpkg{GAS} package.

\section{VaR prediction with the GAS package}

We first briefly review how to make predictions with GAS models using the \CRANpkg{GAS} package. We refer the reader to \citet{ardia_etal.2016} where additional details and examples are presented. The rest of the section is devoted to VaR backtest and model comparison through the DQ test and the QL measure previously introduced.

One--step ahead rolling forecasts applications are straightforward to implement with the \CRANpkg{GAS} package. Specifically, the user just need to specify the model through the \code{UniGASSpec()} function, and then perform rolling predictions with the \code{UniGASRoll()} function. In the \CRANpkg{GAS} package, models are specified through the definition of the conditional distribution assumed for the data, \code{Dist}, and the list of time--varying parameters, \code{GASpar}.

\code{Dist} is a character equal to the label of the distribution. For instance, $\mathcal{SKST}$ is identified as \code{"sstd"}, $\mathcal{ST}$ as \code{"std"} and $\mathcal{N}$ as \code{"norm"}; see Table~1 of \citet{ardia_etal.2016} for the list of distributions and associated labels available in the \CRANpkg{GAS} package.

\code{GASPar} is a \code{list} with named \code{boolean} elements. Entries name are: \code{location}, \code{scale}, \code{skewness} and \code{shape} and indicate whether the associated distribution parameters are time--varying or not. By default we have \code{GASPar = list(location = FALSE, scale = TRUE, skewness = FALSE, shape = FALSE)}, \emph{i.e.}, only volatility is time--varying. For instance, in order to specify the three GAS models: GAS--$\mathcal{SKST}$, GAS--$\mathcal{ST}$ and GAS--$\mathcal{N}$, we need to execute the following lines:

\begin{example}
> library("GAS")
> GASSpec_sstd <- UniGASSpec(Dist = "sstd", GASPar = list(scale = TRUE))
> GASSpec_std  <- UniGASSpec(Dist = "std",  GASPar = list(scale = TRUE))
> GASSpec_norm <- UniGASSpec(Dist = "norm", GASPar = list(scale = TRUE))
\end{example}

\code{UniGASSpec()} delivers an object of the class \code{uGASSpec} which cames with several methods; see \code{help("UniGASSpec")}.

The \code{UniGASRoll()} function accepts an object of the class \code{uGASSpec}, \code{GASSpec}, a \code{numeric} vector for the series of returns, \code{data}, and other arguments such as:

\begin{itemize}
\item the length of the out--of--sample period: \code{ForecastLength},
\item the type of the rolling window used to update the data: \code{RefitWindow},
\item the number of observations within each model re--estimation: \code{RefitEvery},
\end{itemize}

among others; see \code{help("UniGASRoll")}. \code{ForecastLength} and \code{RefitEvery} are \code{numeric} elements while \code{RefitWindow} is a \code{character} equal to \code{"moving"} (the default) for a fixed window scheme or \code{"recursive"} for expanding window. As previously mentioned, in this paper we consider the case \code{RefitWindow = "moving"}. For reproducibility purposes, the \code{dji30ret} data set which consists of the Dow Jones Index 30 Constituents closing value log--returns is included in the \CRANpkg{GAS} package. For our examples, we consider the last $T = 2500$ observations of \code{dji30ret}:

\begin{example}
> data("dji30ret", package = "GAS")
> dji30ret = tail(dji30ret, 2500)
\end{example}

We fix the length of the out--of--sample period to 1000 ($H=1000$), and re--estimate the model parameters each time a new observation arrives. One--step ahead rolling predictions for the first series of returns using the GAS--$\mathcal{N}$ model are then computed as:

\begin{example}
> library("parallel")
> cluster = makeCluster(8)
> H = 1000
> Roll_1N = UniGASRoll(dji30ret[, 1], GASSpec_norm, ForecastLength = H,
                       RefitEvery = 1, cluster = cluster)
\end{example}

where we have also made use of parallel processing through the definition of a \code{cluster} object exploiting the \pkg{parallel} package included in \textsf{R} since version 2.14.0.

The output of \code{UniGASRoll()} is an object of the class \code{uGASRoll} which comes with several methods; see \code{help("UniGASRoll")}. VaR forecasts at confidence level $\alpha = 0.01$ can be computed from \code{Roll\_1N} using the \code{quantile} method, \emph{i.e.}:

\begin{example}
> alpha = 0.01
> VaR_1N = quantile(Roll_1N, probs = alpha)
\end{example}

where \code{VaR\_1N} is a matrix of dimension $1000\times 1$ containing the VaR forecasts.\footnote{The \code{probs} argument in \code{quantile} can also be a \code{numeric} vector of $p$ VaR levels. In this case, \code{VaR\_1N} would be a $1000\times p$ \code{matrix}.} VaR backtest procedures are implemented through the \code{BacktestVaR()} function. This function accepts the following arguments:

\begin{itemize}
\item \code{data}, \code{numeric} containing the out--of--sample data;
\item \code{VaR}, \code{numeric} containing the series of VaR forecasts;
\item \code{alpha}, the VaR confidence level $\alpha$;
\item \code{Lags}, the number of lags used in the DQ test, by default \code{Lags = 4}; see \citet{engle_manganelli.2004}.
\end{itemize}

The function returns a \code{list} with named entries:

\begin{itemize}
\item \code{LRuc}, the test statistic and associated $p$--value for the UC test of \citet{kupiec.1995};
\item \code{LRcc}, the test statistic and associated $p$--value for the CC test of \citet{christoffersen.1998};
\item \code{DQ}, the test statistic and associated $p$--value for the DQ test of \citet{engle_manganelli.2004};
\item \code{Loss}, the average QL and QL series used by \citet{gonzalezriviera_etal.2004},
\item \code{AD}, the mean and max VaR Absolute Deviation (AD) used by \citet{mcaleer_daveiga.2008};
\item \code{AE}, the AE ratio reported in~\eqref{eq:AE}.
\end{itemize}

For instance, in order to compute the VaR backtest measures defined above on the forecast series \code{VaR\_1N}, we use:

\begin{example}
> VaRBacktest_1N = BacktestVaR(data = tail(dji30ret[, 1], H), VaR = VaR_1N, alpha = alpha)
\end{example}

Hence, the DQ test statistic and its associated $p$--value can be extracted as:

\begin{example}
> VaRBacktest_1N$DQ
$stat
        [,1]
[1,] 42.0586

$pvalue
            [,1]
[1,] 1.79043e-07
\end{example}

which, in this case, is against the null of correct model specification for the 1\% VaR level.

Now, if we evaluate VaR forecasts using the GAS--$\mathcal{ST}$ model, we obtain, say, the \code{VaRBacktest\_1ST} object. For GAS--$\mathcal{ST}$ the DQ test reports:

\begin{example}
R> VaRBacktest_1ST$DQ
$stat
        [,1]
[1,] 7.26818

$pvalue
          [,1]
[1,] 0.2967566
\end{example}

The large $p$--value indicates that the null of correct model specification for the 1\% VaR cannot be rejected at the usual levels of signifcance. Models comparison in terms of average QL also favours GAS--$\mathcal{ST}$. Indeed the ratio between the QL of GAS--$\mathcal{ST}$ and GAS--$\mathcal{N}$:

\begin{example}
> round(VaRBacktest_1ST$Loss$Loss / VaRBacktest_1N$Loss$Loss, 2)
[1] 0.89
\end{example}

This indicates that GAS--$\mathcal{ST}$ outperforms GAS--$\mathcal{N}$ by 11\% in terms of average QL.

\section{Empirical application: US Industrial firms}
\label{sec:empirical}

We report now a short empirical application in order to illustrate the benefits of using GAS models for VaR predictions.

We predict the one--step ahead VaR level with the GAS--$\mathcal{N}$, GAS--$\mathcal{ST}$ and GAS--$\mathcal{SKST}$ models using the same configuration of the previous section, \emph{i.e.}, we set, $T = 2500$, $H = 1000$, and \code{RefitEvery = 1}. The analysis considers all the 30 constituents of the Dow Jones Index (DJI) available in the \code{dji30ret} data set previously detailed. The out--of--sample period starts on 14, February 2005, and includes the recent Global Financial Crisis of 2007--2008.

We consider two VaR confidence levels $\alpha = 1\%$ and $\alpha = 5\%$. The code used for this application is available in the GitHub \CRANpkg{GAS} repository: \url{https://github.com/LeopoldoCatania/GAS}.

\begin{figure}[H]
\centering
\includegraphics[width=1.0\textwidth]{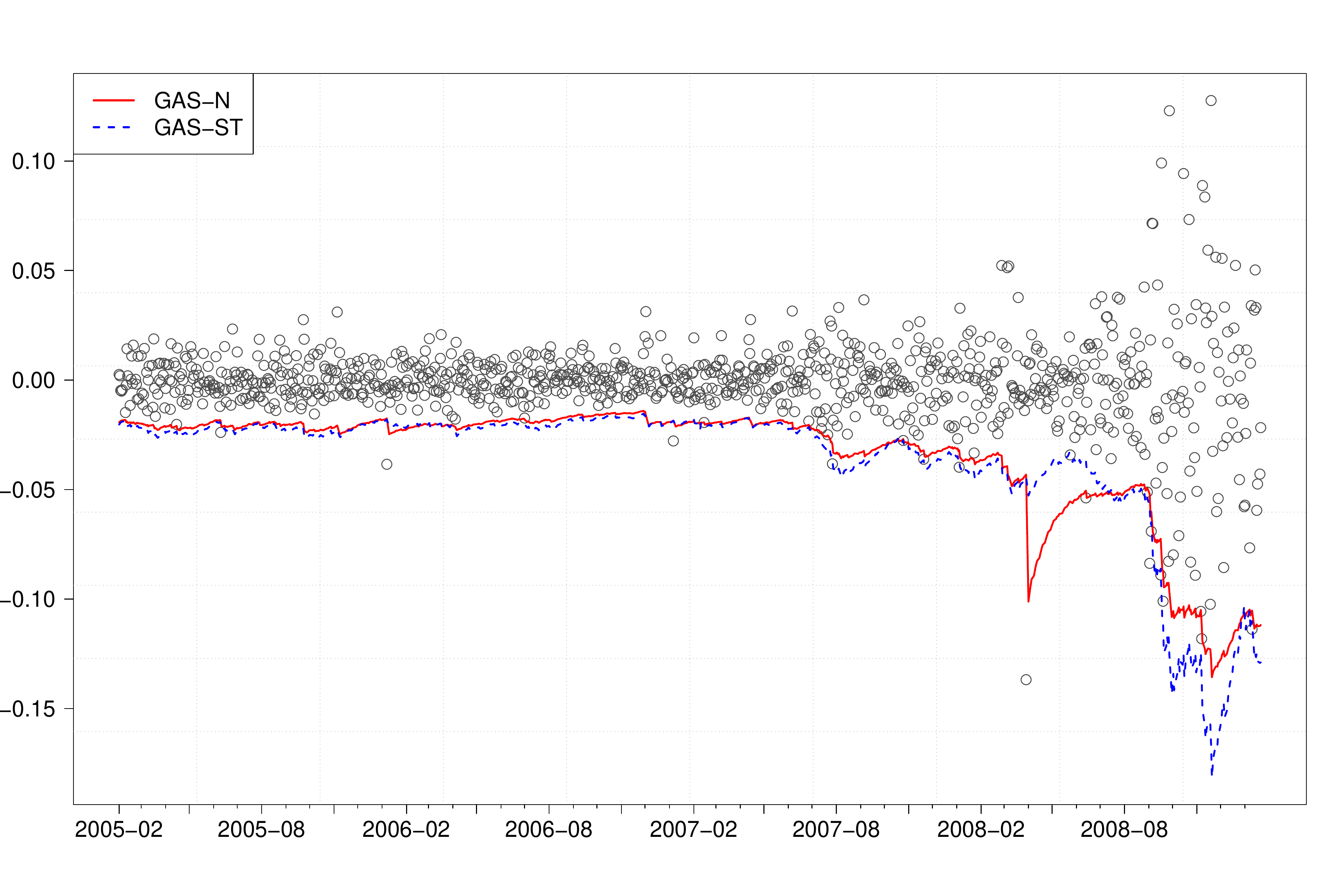}
\caption{\footnotesize{One--step ahead VaR forecasts for General Electric (GE) at the $\alpha = 1\%$ confidence level for the GAS--$\mathcal{N}$ (\emph{solid}) and GAS--$\mathcal{ST}$ (\emph{dotted}) models. Gray points indicate realized log--returns calculated as the differences between the logarithm of two consecutive prices. The forecast period spans from 14, February 2005, to February 3rd, 2009 for a total of 1,000 predictions.}}
\label{fig:VaR}
\end{figure}

Figure~\ref{fig:VaR} depicts the one--step ahead VaR forecasts for General Electric (GE) at the $\alpha = 1\%$ confidence level for the GAS--$\mathcal{N}$ (\emph{solid}) and GAS--$\mathcal{ST}$ (\emph{dotted}) specifications. We clearly see the impact of the recent Global Financial Crisis on the volatility of the series. Indeed, the last part of the figure presents much more variability than the initial one. What is also clearly evident from Figure~\ref{fig:VaR}, is the robustness of the GAS--$\mathcal{ST}$ model to extreme observations compared to GAS--$\mathcal{N}$. Indeed, on 11, April 2008, General Electric reported an unexpected net income drop of 6\%, which in turn translated to a fall of about 12\% of its market value. The signal capture by GAS--$\mathcal{N}$ was that of an abrupt increase in volatility, with the consequence of large VaR level predictions. In contrast, the GAS--$\mathcal{ST}$ model slightly increased the volatility level and continued to predict reasonable VaR levels. What happened is that, GAS--$\mathcal{ST}$ treated the 12\% negative return as realization from the fat--tailed Student--t distribution, hence tapering its impact on the conditional volatility level. On the contrary, GAS--$\mathcal{N}$ treated the negative return as a realization from the Normal distribution, which is a clear signal of increase of volatility.

Table~\ref{tab:DQ} reports the DQ test $p$--values for the three model specifications and the two VaR confidence levels.

\begin{table}[H]
\centering
\scalebox{0.9}{
\begin{tabular}{lcccccc}
\toprule
& \multicolumn{3}{c}{$\alpha = 1\%$} & \multicolumn{3}{c}{$\alpha = 5\%$}\\
\cmidrule(lr){1-1}\cmidrule(lr){2-4}\cmidrule(lr){5-7}
Asset & GAS--$\mathcal{N}$ & GAS--$\mathcal{ST}$ & GAS--$\mathcal{SKST}$ & GAS--$\mathcal{N}$ & GAS--$\mathcal{ST}$ & GAS--$\mathcal{SKST}$\\
\cmidrule(lr){1-1}\cmidrule(lr){2-4}\cmidrule(lr){5-7}
AA & \grb{0.00} & 0.29 & 0.29 & \grb{0.00} & \grb{0.00} & \grb{0.00} \\
AIG & \grb{0.00} & \grb{0.00} & \grb{0.00} & \grb{0.00} & \grb{0.00} & \grb{0.00} \\
AXP & 0.08 & 0.99 & 0.67 & 0.17 & 0.18 & 0.08 \\
BA & 0.21 & 0.92 & 0.20 & 0.09 & 0.36 & 0.40 \\
BAC & \grb{0.00} & \grb{0.00} & 0.08 & \grb{0.00} & 0.02 & 0.06 \\
C & \grb{0.00} & \grb{0.00} & \grb{0.00} & \grb{0.01} & \grb{0.00} & \grb{0.01} \\
CAT & \grb{0.00} & 0.15 & 0.09 & \grb{0.00} & 0.51 & 0.48 \\
CVX & \grb{0.00} & 0.45 & 0.20 & 0.02 & 0.02 & 0.02 \\
DD & 0.03 & 0.31 & 0.24 & 0.26 & 0.22 & 0.11 \\
DIS & \grb{0.00} & 0.98 & 0.23 & \grb{0.00} & 0.28 & 0.12 \\
GE & \grb{0.00} & 0.21 & 0.25 & 0.17 & 0.33 & 0.03 \\
GM & \grb{0.00} & 0.04 & 0.01 & 0.02 & 0.19 & 0.11 \\
HD & 0.02 & \grb{0.01} & 0.06 & 0.37 & 0.89 & 0.67 \\
HPQ & \grb{0.00} & 0.01 & 0.02 & \grb{0.00} & 0.06 & 0.17 \\
IBM & \grb{0.00} & 0.02 & 0.02 & \grb{0.00} & 0.02 & 0.02 \\
INTC & 0.02 & 0.04 & 0.05 & 0.02 & 0.04 & 0.05 \\
JNJ & 1.00 & 1.00 & 0.96 & 0.01 & \grb{0.00} & \grb{0.00} \\
JPM & 0.24 & 0.93 & 0.27 & 0.29 & 0.42 & 0.17 \\
KO & 0.06 & 0.03 & 0.20 & \grb{0.00} & 0.12 & 0.22 \\
MCD & 0.81 & 0.98 & 1.00 & 0.27 & 0.18 & 0.40 \\
MMM & \grb{0.00} & 0.29 & \grb{0.00} & \grb{0.00} & 0.13 & 0.07 \\
MRK & \grb{0.00} & 0.06 & 0.06 & \grb{0.00} & 0.08 & 0.03 \\
MSFT & 0.07 & 0.20 & 0.31 & 0.30 & 0.36 & 0.44 \\
PFE & 0.13 & 0.28 & 0.22 & 0.09 & 0.78 & 0.76 \\
PG & \grb{0.00} & 0.12 & 0.08 & 0.16 & 0.10 & 0.19 \\
T & 0.26 & 0.27 & 0.31 & 0.02 & \grb{0.01} & 0.04 \\
UTX & \grb{0.00} & 0.31 & 0.04 & 0.17 & 0.82 & 0.93 \\
VZ & \grb{0.01} & 0.99 & 0.98 & 0.89 & 0.81 & 0.90 \\
WMT & \grb{0.01} & \grb{0.00} & \grb{0.01} & 0.83 & 0.21 & 0.10 \\
XOM & \grb{0.00} & \grb{0.00} & \grb{0.00} & 0.08 & 0.02 & 0.04 \\
\bottomrule
\end{tabular}
}
\caption{DQ test statistic $p$--values for DJI constituents one--step ahead VaR forecasts at the two confidence levels $\alpha = 1\%$ and $\alpha = 5\%$. Under the null hypothesis we have correct model specification for the chosen quantile level. Light gray cells indicate $p$--values lower than 1\%. The out--of--sample period spans from 14, February 2005, to February 3rd, 2009 for a total of 1,000 observations.}
\label{tab:DQ}
\end{table}

Under the null hypothesis we have correct model specification for the $\alpha$--quantile level. Our results suggest that, GAS--$\mathcal{ST}$ and GAS--$\mathcal{SKST}$ perform similar in terms of correct unconditional and conditional coverage. Hence, the inclusion of skewness does not seems to increase the performance of VaR predictions for the considered series. Indeed, sometimes results are even worst for GAS--$\mathcal{SKST}$ compared to GAS--$\mathcal{ST}$, indicating that the estimation error for the additional skewness parameter could worsen VaR predictions. This is the case for example for General Electric (GE) when GAS--$\mathcal{SKST}$ reject the null both for $\alpha = 1\%$ and $\alpha = 5\%$, whereas GAS--$\mathcal{ST}$ does never reject.

GAS--$N$ is suboptimal with respect to GAS--$\mathcal{ST}$ and GAS--$\mathcal{SKST}$. This result is somehow expected since GAS--$\mathcal{ST}$ and GAS--$\mathcal{SKST}$ exhibit excess kurtosis and deliver more robust updates for the volatility parameter than GAS--$\mathcal{N}$.

\begin{table}[H]
\centering
\scalebox{0.9}{
\begin{tabular}{lcccccc}
\toprule
& \multicolumn{3}{c}{$\alpha = 1\%$} & \multicolumn{3}{c}{$\alpha = 5\%$}\\
\cmidrule(lr){1-1}\cmidrule(lr){2-4}\cmidrule(lr){5-7}
Asset & GAS--$\mathcal{N}$ & GAS--$\mathcal{ST}$ & GAS--$\mathcal{SKST}$ & GAS--$\mathcal{N}$ & GAS--$\mathcal{ST}$ & GAS--$\mathcal{SKST}$\\
\cmidrule(lr){1-1}\cmidrule(lr){2-4}\cmidrule(lr){5-7}
AA & 1.00 & 0.92 & 0.93 & 1.00 & 0.98 & 0.98 \\
AIG & 1.00 & 0.98 & \grb{1.00} & 1.00 & \grb{1.02} & \grb{1.02} \\
AXP & 1.00 & 0.93 & 0.96 & 1.00 & 0.99 & 1.00 \\
BA & 1.00 & 0.99 & 0.99 & 1.00 & 0.99 & 0.99 \\
BAC & 1.00 & 0.83 & 0.85 & 1.00 & \grb{1.01} & \grb{1.02} \\
C & 1.00 & 0.93 & 0.94 & 1.00 & \grb{1.01} & \grb{1.01} \\
CAT & 1.00 & 0.84 & 0.85 & 1.00 & 0.94 & 0.94 \\
CVX & 1.00 & 1.00 & \grb{1.00} & 1.00 & \grb{1.01} & \grb{1.01} \\
DD & 1.00 & 0.95 & 0.96 & 1.00 & 0.98 & 0.98 \\
DIS & 1.00 & 0.97 & 0.97 & 1.00 & \grb{1.01} & \grb{1.01} \\
GE & 1.00 & 0.94 & 0.97 & 1.00 & 0.97 & 0.98 \\
GM & 1.00 & 0.89 & 0.91 & 1.00 & 0.95 & 0.96 \\
HD & 1.00 & \grb{1.02} & \grb{1.02} & 1.00 & \grb{1.00} & \grb{1.00} \\
HPQ & 1.00 & 0.96 & 0.97 & 1.00 & 0.95 & 0.95 \\
IBM & 1.00 & 1.00 & 1.00 & 1.00 & 0.94 & 0.94 \\
INTC & 1.00 & 0.96 & 0.95 & 1.00 & 0.95 & 0.95 \\
JNJ & 1.00 & 0.99 & \grb{1.00} & 1.00 & 0.99 & 1.00 \\
JPM & 1.00 & 0.92 & 0.94 & 1.00 & 0.99 & 0.99 \\
KO & 1.00 & 0.99 & 0.98 & 1.00 & 0.97 & 0.97 \\
MCD & 1.00 & \grb{1.00} & \grb{1.00} & 1.00 & 0.99 & 0.99 \\
MMM & 1.00 & 0.85 & 0.86 & 1.00 & 0.92 & 0.91 \\
MRK & 1.00 & 0.84 & 0.84 & 1.00 & 0.89 & 0.89 \\
MSFT & 1.00 & 0.88 & 0.89 & 1.00 & 0.93 & 0.92 \\
PFE & 1.00 & 0.85 & 0.85 & 1.00 & 0.92 & 0.92 \\
PG & 1.00 & 0.88 & 0.89 & 1.00 & 0.97 & 0.97 \\
T & 1.00 & 0.97 & 0.96 & 1.00 & \grb{1.01} & \grb{1.01} \\
UTX & 1.00 & 0.98 & 0.99 & 1.00 & \grb{1.00} & 1.00 \\
VZ & 1.00 & 0.91 & 0.90 & 1.00 & 0.98 & 0.98 \\
WMT & 1.00 & 0.95 & 0.96 & 1.00 & 0.99 & 0.99 \\
XOM & 1.00 & \grb{1.02} & 0.99 & 1.00 & \grb{1.02} & \grb{1.01} \\
\bottomrule
\end{tabular}
}
\caption{QL ratios for GAS--$\mathcal{ST}$ and GAS--$\mathcal{SKST}$ with respect to GAS--$\mathcal{N}$ for the two VaR confidence levels $\alpha = 1\%$ and $\alpha = 5\%$. Values greater than 1 indicate outperformance of GAS--$\mathcal{N}$ and vice versa. Light gray cells indicate ratios higher than 1, \emph{i.e.}, when the GAS--$\mathcal{N}$ performs better. The out--of--sample period spans from 14, February 2005, to February 3rd, 2009 for a total of 1,000 observations. }
\label{tab:QL}
\end{table}

Table~\ref{tab:QL} reports the ratios between the average QL of the considered models over the one delivered by GAS--$\mathcal{N}$ for $\alpha = 1\%$ and $\alpha = 5\%$. Values greater than one indicate outperformance of GAS--$\mathcal{N}$, and vice versa. Consistently with the DQ test results, we find that GAS--$\mathcal{ST}$ and GAS--$\mathcal{SKST}$ perform similar and are preferred to GAS--$\mathcal{N}$. Indeed, GAS--$\mathcal{ST}$ outperforms GAS--$\mathcal{N}$ up to 16\% for $\alpha = 1\%$ and 11\% for $\alpha = 5\%$ in terms of average QL.

\section{Conclusion}
\label{sec:conclusion}

This article detailed how to use GAS models for VaR predictions in \code{R} using the \CRANpkg{GAS} package of \citet{GAS}. We briefly review the four steps
 that compose a typical VaR analysis which are: (i) models specifications, (ii) VaR predictions, (iii) backtesting and (iv) model comparison. We illustrate the use of the GAS package for VaR anlaysis using DJI constituents returns series. Our results indicate that the use of a fat--tailed conditional distribution is required by the considered series of returns.

\address{David Ardia \\
Institute of Financial Analysis\\
University of Neuch\^atel, Switzerland\\
and\\
Department of Finance, Insurance and Real Estate\\
Laval University, Canada\\}
\email{david.ardia@unine.ch}

\address{Kris Boudt\\
Vrije Universiteit Brussel, Belgium\\
and\\
Vrije Universiteit Amsterdam, The Netherlands\\}
\email{kris.boudt@vub.ac.be}

\address{Leopoldo Catania (corresponding author)\\
Department of Economics and Finance\\
University of Rome, \qmo Tor Vergata\qmc, Italy\\}
\email{leopoldo.catania@uniroma2.it}
\end{article}

\end{document}